pdfoutput=1 

\documentclass{vgtc}                          

\usepackage{mathptmx}
\usepackage{graphicx}
\usepackage{times}
\usepackage{caption}
\usepackage{subcaption}
\usepackage{rotating}

\onlineid{0}

\vgtccategory{Research}

\vgtcinsertpkg

\title{A novel display for situational awareness at a network operations center}

\author{Andrea Brennen, David Danico, Raul Harnasch, Maureen Hunter, \vspace{0.1 cm}\\
               Richard Larkin, Jeremy Mineweaser, Kevin Nam, David O'Gwynn, \vspace{0.1 cm}\\
                Harry Phan, Alexia Schulz, Michael Snyder, Diane Staheli, Tamara Yu
                \thanks{e-mail: [abrennen, danico,  Raul.Harnasch, Maureen.Hunter, 
                                             larkin, jlm, Kevin.Nam, dogwynn, 
                                             Harry.Phan, Alexia.Schulz, Michael.Snyder, Diane.Staheli, tamara]@ll.mit.edu} %
                 \vspace{0.4 cm} \\ %
                 \scriptsize MIT Lincoln Laboratory }
\abstract{As modern industry shifts toward significant globalization, robust and adaptable network capability is increasingly vital to 
the success of business enterprises.   Large quantities of information must be distilled and presented in a single integrated picture
in order to maintain the health, security and performance of global networks.  We present a design for a network 
situational awareness display that visually aggregates large quantities of data, identifies problems in a network, assesses their impact 
on critical company mission areas and clarifies the utilization of resources.  This display facilitates the prioritization of network problems as they 
arise by explicitly depicting how problems interrelate.  It also serves to coordinate mitigation strategies with members of a team. 
} 

\begin{document}


\firstsection{Introduction}

\maketitle

The Visual Analytics for Science and Technology Challenge 2013 commissions an innovative graphical design to support situation awareness 
for large-scale computer networks.  The scenario depicts a company, Big Enterprise, which requires a big board display
to be used in their Network Operations Center (NOC) for visually assessing the health, security and 
performance of the network at a glance. This interactive display must be designed to provide a concise visual representation of the state of 
the Big Enterprise network, comprised of thousands of systems in major cities spread over four continents.   

We present a design that will allow the network operations manager to quickly identify and classify problems in the network, 
assess their impact on critical mission areas, prioritize the utilization of resources, and coordinate mitigation strategies with other members of the team.  
The display is interactive; users can augment the comprehensive network overview with additional layers of 
information that highlight particular problem areas.   The NOC manager can use the 
highlighted overlays to feature company mission areas, geographical regions, or functional categories in the network (e.g. all facilities affected by an outdated patch).   Multiple 
highlights are simultaneously possible, creating a visual representation of the impact of network problems on several domains, as well as how the domains intersect 
one another.  This display will significantly enhance the ability of the NOC manager to quickly prioritize multiple problems, and coordinate a strategy with her team.  

\section{Display Design}
The Big Enterprise Big Board display is an interactive visual display designed to provide a concise visual representation of the state of the company network.  The 
goal of the display is to facilitate the prioritization of network problems as they arise, by explicitly depicting how problems interrelate.  The board depicts how network
problems are geographically or functionally distributed, and illustrates how they impact critical enterprise mission areas.  Two images of the board can be seen in 
Figure \ref{fig:both}.

\subsection{Layout}
The layout of the Big Enterprise Big Board is designed to aggregate information to provide a high level overview of the network, while simultaneously drawing
attention to potential network problems.  The underlying philosophy of the design is that properly functioning portions of the network are depicted with representative 
aggregates only, whereas systems that are not functioning properly are highlighted in a more individual way.   Visual alerts quickly draw users' attention to problems arising in the network.  The layout has four main sections, shown in Figure \ref{fig:both}; the central panel, the warning menu to the right, the mission
maps and functional queries for dynamic overlay to the left, and the net flow pipes at the bottom.  The network name appears in the top left of the display, and the time at the top right. 

\subsubsection{The central panel}
The central portion of the display shows the aggregated network resources of the company.  These are broken down into a number of zones that organize the assets of the  network according to broad functional categories.  In the case of the hypothetical company Big Enterprise, there are five zones shown in Figure \ref{fig:both}: Virtual 
Private Network users, office locations (which are organized geographically), core services, the data center and the extranet.  Within each zone, the resources are further 
subdivided into more specific functional categories, for example, the core services in Figure \ref{fig:both} include network defense, E-mail, and Domain Name Servers.  Different 
shading schemas within these subdivisions can be used to further specify details of the network hierarchy.   The zones and sub-zones are customizable and should reflect the 
functional breakdown of whichever company is using the display.

\subsubsection{Icons and warnings}
\begin{sidewaysfigure*}
\centering
\begin{subfigure}{1\textheight}
  \centering
  \includegraphics[width=9in]{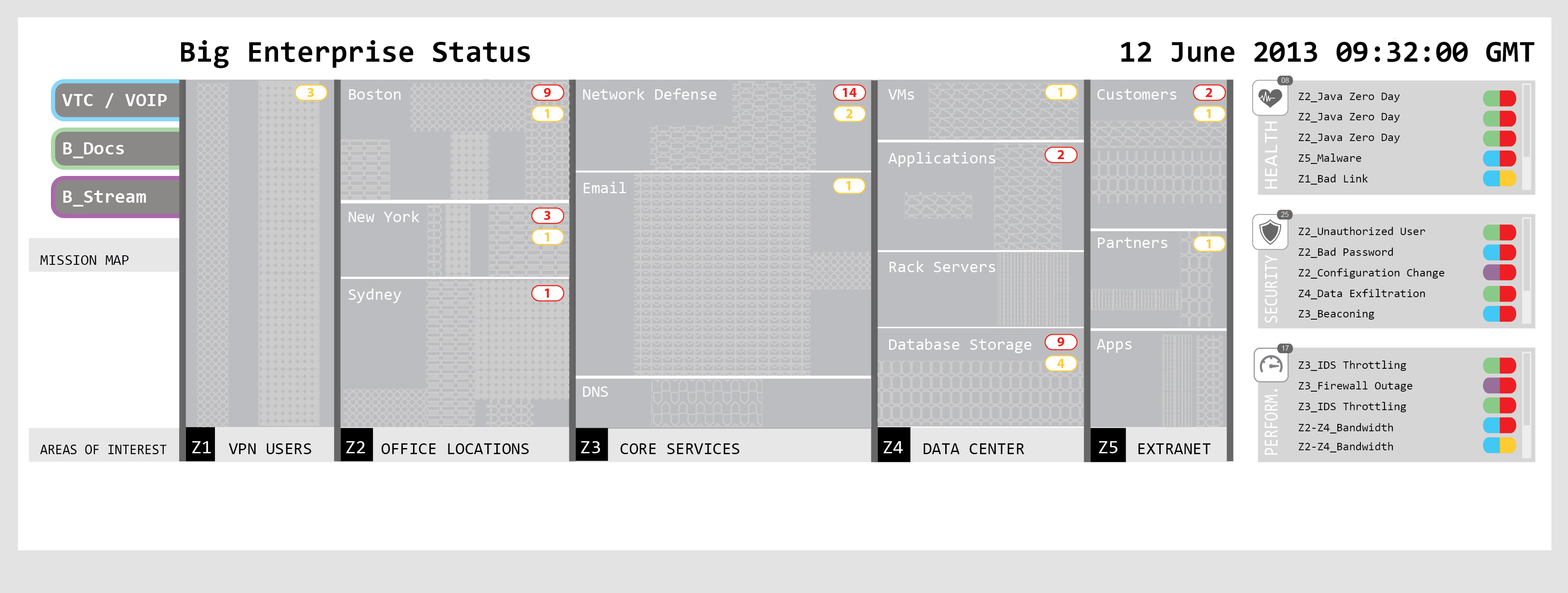}
  \caption{The Big Board display with no active highlights or layers.}
  \label{fig:bigboard}
\end{subfigure}%

\vspace{0.5 cm}
\begin{subfigure}{1\textheight}
  \centering
  \includegraphics[width=9in]{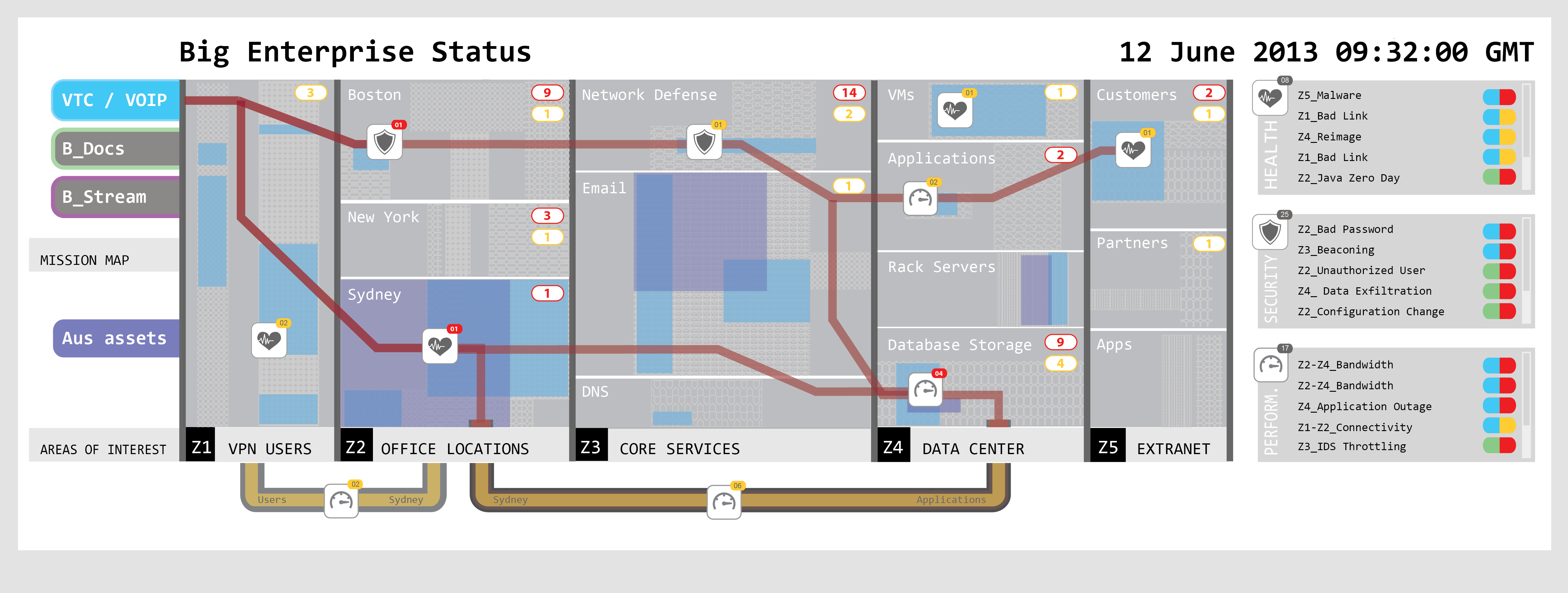}
  \caption{The Big Board, with one mission tab activated, and one functional search dynamically overlaid.}
  \label{fig:mission}
\end{subfigure}
\caption{The Big Enterprise Big Board display.  {\bf (a)} The display is depicted in its simplest view.  {\bf (b)} The Big Enterprise big board display is shown here with one mission area highlighted: Video Teleconference and Voice over IP. The image also shows the results of one active functional query; assets physically located in Australia are highlighted in purple. }
\label{fig:both}
\end{sidewaysfigure*}

When specific problems arise, pipes and icons appear on the central panel of the board, with details about the affected systems displayed in the warning menus to the right of the 
central panel.  Alerts are divided into three main categories; health, performance and security.  New problems are triaged and added to the board with a red 
flag, indicating that these are problems that need to be assigned for resolution.  When problems are tasked to an individual or a team, the red flag is changed to yellow, indicating 
that the problem is being actively addressed but is not yet resolved.  The aggregation of multiple assets into different grey zones significantly simplifies the view of healthy 
systems, while the pop-up notifications draw attention to portions of the network that currently require action.  Because the network is so large, it is also necessary to aggregate 
the alerts.  The aggregated flags for each sub-zone can be seen in red and yellow ovals; for example in Figure \ref{fig:both}, the Boston sub-zone has nine red alerts and 
one 
yellow one.  Figure \ref{fig:mission} also shows alert badges that indicate individual unaggregated problems in the network, such as the heart shaped badge in Sydney.  
There are three possible icons used for these alert badges: a heart for health, a shield for security, and a speedometer for performance.  These individual alert 
badges are discussed in more detail in section \ref{sec:mission map}.

The warning menus on the right hand side provide more detail about the aggregated alerts in the central panel.  When there are more alerts than can fit into the display, the 
alerts in each category scroll slowly by, so that within a few moments, any alert in any zone can be identified.  Each item in the warning menu is accompanied by a pill shaped 
capsule to the right.  The right portion of the capsule indicates whether the alert has been tasked out for resolution (yellow), or is still unassigned (red).  
The left hand portion of the 
capsule is color coded to indicate which mission area is most significantly affected by this particular outage or problem.  This 
mission context is discussed further in section \ref{sec:mission map}.

With thousands of machines distributed globally, it is not useful to depict all network connections and their available bandwidth.  Nonetheless, the display is designed to 
indicate connectivity issues.  Pipes at the bottom of the display are used to visualize bandwidth problems between elements in
different zones.  The pipe is labeled at each end to indicate which specific zone subdivisions are experiencing problems, for example, in Figure \ref{fig:mission} a pipe shows 
a connectivity issue between VPN users and the Sydney offices.  Pipe alerts are also included in the warning menu on the right.  The outer layer of the pipe is a 
qualitative visual indicator of the available bandwidth, while the inner pipe shows the current capacity, updated in real time.  The color coding is consistent with the health, 
security, and performance icons; the pipes are colored red if the connectivity problem has been noted but not assigned to anyone, and turn yellow once a service ticket has been 
issued.  The pipes disappear when the problem is resolved.  

\subsubsection{The mission map} \label{sec:mission map}
The most powerful feature of this design is the ability to illustrate visually how problems in the network impact specific company missions.  An enterprise
mission area is a key business 
objective, the loss of which would damage the company holdings or reputation.  The Big Enterprise scenario has three: VTC/VOIP internet voice capability, B\_Docs in analogy 
with Google Docs, and B\_Stream akin to Rhapsody or iTunes.   
The company missions are displayed in color coded tabs at the top left of the display.  These tabs can be 
active or inactive.  In Figure \ref{fig:mission}, the VTC/VOIP mission is activated, while the B\_docs and B\_stream are inactive.  Activating the tab changes it to a solid 
color, and results in several additional layers being added to the display.  
First, the network assets that directly impact this particular mission are highlighted on the central panel, color 
coordinated with the tab and the warning menu capsules.  This is seen in Figure \ref{fig:mission} in the blue shaded areas on the central panel.   Second, a map connecting 
unresolved (red) network problems  is drawn with a red strip.  The strip connects larger icons, whose symbols classify the issue as relevant to security, health or performance.  
These icons represent individual (non-aggregated) problems that specifically impact this mission area. All network problems pertaining to this mission are displayed individually, 
but the map brings focus to network anomalies that must be tasked in order to keep this mission functional. 

Clicking the tab also simplifies the display and helps the network manager to sort the information based on its mission impact.   
Activating the tab causes unrelated pipes at the bottom to 
temporarily disappear, and changes the ordering of the items in the warning menu to highlight the warnings specific to this particular mission. 

\subsubsection{Dynamic overlay}
The Big Board design also incorporates an adaptable feature that helps the network manager augment the situational awareness of her team.  This feature implements
a dynamic 
overlay of areas of interest.  Functionally, it is a search tool with the flexibility to highlight either geographical or functional categories.  
One example, a search for assets in Australia, is shown in Figure 
\ref{fig:mission} in purple.  The search is intended to be flexible enough to accommodate a variety of queries.  Some examples might be ``hosts with unpatched Java,"  ``Beaconing 
activity," ``hosts bypassing the proxy," or  ``hosts connecting to 194.220.1.0."   Activating a functional search causes 
a colored tab to appear in the lower left of the display below the mission map tabs, and color coordinated highlighting to be overlaid on the central panel.  This highlighting
intersects alert icons and the assets in each zone, conveying the dependency of company assets on the functional category being highlighted.  There can be as many as eight 
color coded areas of interest.  These can be saved and toggled on and off in much the same way as the mission maps.  
\subsection{Interaction channels}
The Big Board display is controlled primarily though a tablet device such as an iPad belonging to the NOC manager.  However, the app can be used by any team
member to check out a local copy of the Big Board on a mobile device or computer.  The local copies can be interacted with as a mini version of the actual Big Board.   
This will allow individual team members to explore network problem interdependencies, helping them to triage problems.  Red alerts can be added to the Big Board at any time 
from any of the local copies.  Red alerts are coordinated with the ticket system, so that they turn yellow when a ticket has been accepted by a team member for resolution. 
Any local copy of the Big Board can access notes about the particular network problem that have been made in the ticket system.  Once the tickets are resolved, the yellow alerts 
can be removed using the local copy of the application. 

The NOC manager is responsible for the view presented on the Big Board, 
namely which missions are highlighted at this time, and which dynamic functional searches are visible.   
Functional searches are initiated through a search tool deployed on her mobile platform, and the highlighting can be toggled on and off as she deems necessary to communicate 
her priorities with her team. 
\section{Summary}
The VAST Mini-Challenge 2013 commissioned the design of a Big Board display for a hypothetical multinational company with a globalized network, Big Enterprise.  
In this paper we have presented a visual display that concisely summarizes the health, security and performance of the Big Enterprise network.  The display illustrates how network problems are geographically or functionally distributed, and how they affect critical mission areas within Big Enterprise.  The principle features of the display are 
listed below.  
\begin{itemize} 
\item Network assets are divided into functional zones, which are further divided by geography or by category.
\item Correctly functioning assets are aggregated while network problems are organized and displayed with clarity.
\item Color coding clearly differentiates unaddressed network problems from those that are currently tasked out.
\item Mission highlighting illustrates the intersection space of mission critical assets with current network problems. 
\item Functional search queries dynamically add a highlighting layer that show the intersection of that function with company assets in each zone, and with mission areas. 
\end{itemize}
This display significantly increases network situation awareness in a global context, and provides critical insight into the optimal way to prioritize problems and allocate resources and labor. 

\vspace{0.2 cm}
{\bf \tiny This work is sponsored by the Assistant Secretary of Defense for Research \& Engineering under Air Force Contract \#FA8721-05-C-0002.  Opinions, interpretations, conclusions and recommendations are those of the authors and are not necessarily endorsed by the United States Government.\par}

\newpage

\end{document}